\documentclass[%
 reprint,
superscriptaddress,
 amsmath,amssymb,
 aps,
 prl,
]{revtex4-2}

\usepackage{graphicx}
\usepackage{dcolumn}
\usepackage{bm}
\usepackage{subcaption,amsmath,gensymb,esvect,xcolor,xr,multirow,commath}
\usepackage[colorlinks=true,citecolor=blue]{hyperref}
\usepackage[normalem]{ulem}

\begin{document}

\title{Monolayer MnX and Janus XMnY (X, Y= S, Se, Te): A New Family of 2D Antiferromagnetic Semiconductors}

\author{Shahid Sattar}
 \email{shahid.sattar@lnu.se}
\author{M. F. Islam}
 \author{C. M. Canali}
 \affiliation{Department of Physics and Electrical Engineering, Linnaeus University, SE-39231 Kalmar, Sweden}
 \date{\today}
\date{\today}

\begin{abstract}

We present first-principles results on the structural, electronic, and magnetic properties of a new family of two-dimensional antiferromagnetic (AFM) manganese chalcogenides, namely monolayer MnX and Janus XMnY (X, Y= S, Se, Te), among which monolayer MnSe was recently synthesized in experiments [\href{https://pubs.acs.org/doi/abs/10.1021/acsnano.1c05532}{ACS Nano 15 (8),13794 (2021)}]. By carrying out calculations of the phonon dispersion and \textit{ab-initio} molecular dynamics simulations, we first confirmed that these systems, characterized by an unconventional strongly-coupled-bilayer atomic structure (consisting of Mn atoms buckled to chalcogens forming top and bottom ferromagnetic (FM) planes with antiparallel spin orientation) are dynamically and thermally stable. The analysis of the magnetic properties shows that these materials have robust AFM order, retaining a much lower energy than the FM state even under strain. Our electronic structure calculations reveal that pristine MnX and their Janus counterparts are indirect-gap semiconductors, covering a wide energy range and displaying tunable band gaps by the application of biaxial tensile and compressive strain. Interestingly, owing to the absence of inversion and time-reversal symmetry, and the presence of an asymmetrical potential in the out-of-plane direction, Janus XMnY become spin-split gapped systems, presenting a rich physics yet to be explored. Our findings provide  novel insights in this physics, and highlight the potential for these two-dimensional manganese chalcogenides in AFM spintronics. 

\end{abstract}

\maketitle

\section{Introduction}

The experimental realization of two-dimensional (2D) magnetic crystals has accelerated and enhanced the possibility of controlling and harnessing the spin degree of freedom for spintronics applications. Transition metal (TM) atoms with 3d open shells and spin-orbit coupling (SOC) induce large magnetocrystalline anisotropy or magnetic dipolar interactions, which are essential for intrinsic FM order in chromium trihalides (CrH$_3$ (H= Cl, Br, I) \cite{klein2019enhancement,ghazaryan2018magnon,huang2017layer}), TM germanium tellurides (Cr$_2$Ge$_2$Te$_6$ \cite{gong2017discovery}, Fe$_3$GeTe$_2$ \cite{deng2018gate}, Fe$_4$GeTe$_2$ \cite{seo2020nearly}) and several other 2D magnets \cite{bonilla2018strong,wu2021strong,pham2020tunable,chu2019sub}. Beside their standalone intrinsic magnetism, another peculiar feature is the possibility of building stacked heterostructures made up of ultrathin FMs and other materials, enabling spin-orbit torque (SOT) switching \cite{shin2021spin}, valley polarization \cite{zhong2017van,seyler2018valley}, skyrmionics \cite{wu2020neel}, and other novel phenomena.    

On the other side, antiferromagnetic (AFM) materials have recently triggered great interest and attention owing to their integration in building non-volatile memory devices \cite{jungwirth2016antiferromagnetic}, neuromorphic computing \cite{kurenkov2020neuromorphic}, and THz scale information processing \cite{baltz2018antiferromagnetic}. Having finite internal magnetic moments (with anti-parallel orientation) and a vanishing net magnetization, AFMs are unresponsive and externally invisible to an applied external magnetic field and do not generate stray fringing fields like FMs. The possibility to achieve current-induced local spin-polarization at non-centrosymmetric lattice sites of an AFM with opposite magnetic moments provides electrical control of spin for read/write operations \cite{vzelezny2014relativistic,bodnar2018writing}. Broken structural inversion symmetry creating staggered current-induced non-equilibrium fields are deemed crucial to achieve such functionality \cite{jungwirth2018multiple}. Moreover, AFM/FM heterostructures show analogue-like SOT induced magnetization switching in FMs without external magnetic field (i.e., field-free switching) and mimic synapses of the brain \cite{fukami2016magnetization,oh2016field,borders2016analogue}. Such device architectures offer multilevel switching which allows functionalities of logic and memory within the same bit cell by adjusting synaptic weights of SOT operation \cite{grollier2020neuromorphic}. Another interesting feature of AFMs is their ultrafast information processing (in THz regime) compared to their FM counterpart which demands excessive energies to operate beyond GHz scale \cite{Garello2014a,olejnik2017antiferromagnetic,liu2022nanoplasmonic}. While these features make AFMs highly interesting and appealing for the next-generation of electronic and spintronics devices, studies of 2D AFMs with broken inversion symmetry, large SOC strength, and semiconducting behavior are rather scarce and demands attention \cite{zhang2019magnetism}. Such ultrathin magnets are also suitable for optical control and manipulation of their spin degree of freedom \cite{nvemec2018antiferromagnetic}. 

Bulk manganese chalcogenides in either rock-salt (NaCl) or hexagonal (NiAs) crystal structure and AFM ordering have been known for decades \cite{pollard1983magnetic,komatsubara1963magnetic}. However, their 2D counterparts has gained attention only recently owing to the discovery of room-temperature intrinsic FM arrangement in a single-layer MnSe$_2$ \cite{o2018room} and ultrahigh photonic responsivity in non-layered $\alpha$-MnTe \cite{li2020chemical}. Alongside several examples \cite{NovoselovScience2016,wang2015highly,wang2015monolayer}, it clearly demonstrates that materials in 2D limit can host unusual phenomenon otherwise absent in their bulk-analogues. Following to it, monolayer MnSe with unusual atomic structure and AFM ordering has been recently synthesized in experiments \cite{aapro2021synthesis}. Interestingly, owing to the presence of top and bottom FM planes coupled together via AFM coupling, this new material has opened up exciting avenues for AFM spintronics. Here, the presence of inversion asymmetric lattice sites of top and bottom Mn atoms, together with large SOC strength and robust AFM coupling, offer countless possibilities discussed in Ref. \cite{vzelezny2014relativistic,bodnar2018writing,jungwirth2018multiple}. In addition, Janus AFM 2D systems, elements of this class characterized by the two chalcogen (X= S, Se, Te) atoms being dissimilar, can provide additional control of the electron and spin degrees of freedom. Several 2D Janus materials have been recently explored and experimentally synthesized (such as SMoSe \cite{zhang2017janus,qin2022reaching}, SPtSe \cite{sant2020synthesis} etc). However, so far the focus has been mainly on non-magnetic systems with plenty of space left for magnetic materials.

In this paper we employ first-principles calculations to investigate the structural, electronic, and magnetic properties of the entire family of 2D AFM manganese chalcogenides, namely monolayer MnX and Janus XMnY (X, Y= S, Se, Te). After initially confirming the thermal and dynamical stability of all members, we demonstrate the presence of robust AFM ordering which remains intact under biaxial tensile and compressive strain. Thanks to the presence of asymmetrical lattice sites and an indirect band gap tunable over a large energy range, we predict that these magnetic chalcogenides provide exciting prospects for non-volatile memory applications and opto-spintronics.             

\begin{figure}[!ht]
\includegraphics[width=0.47\textwidth]{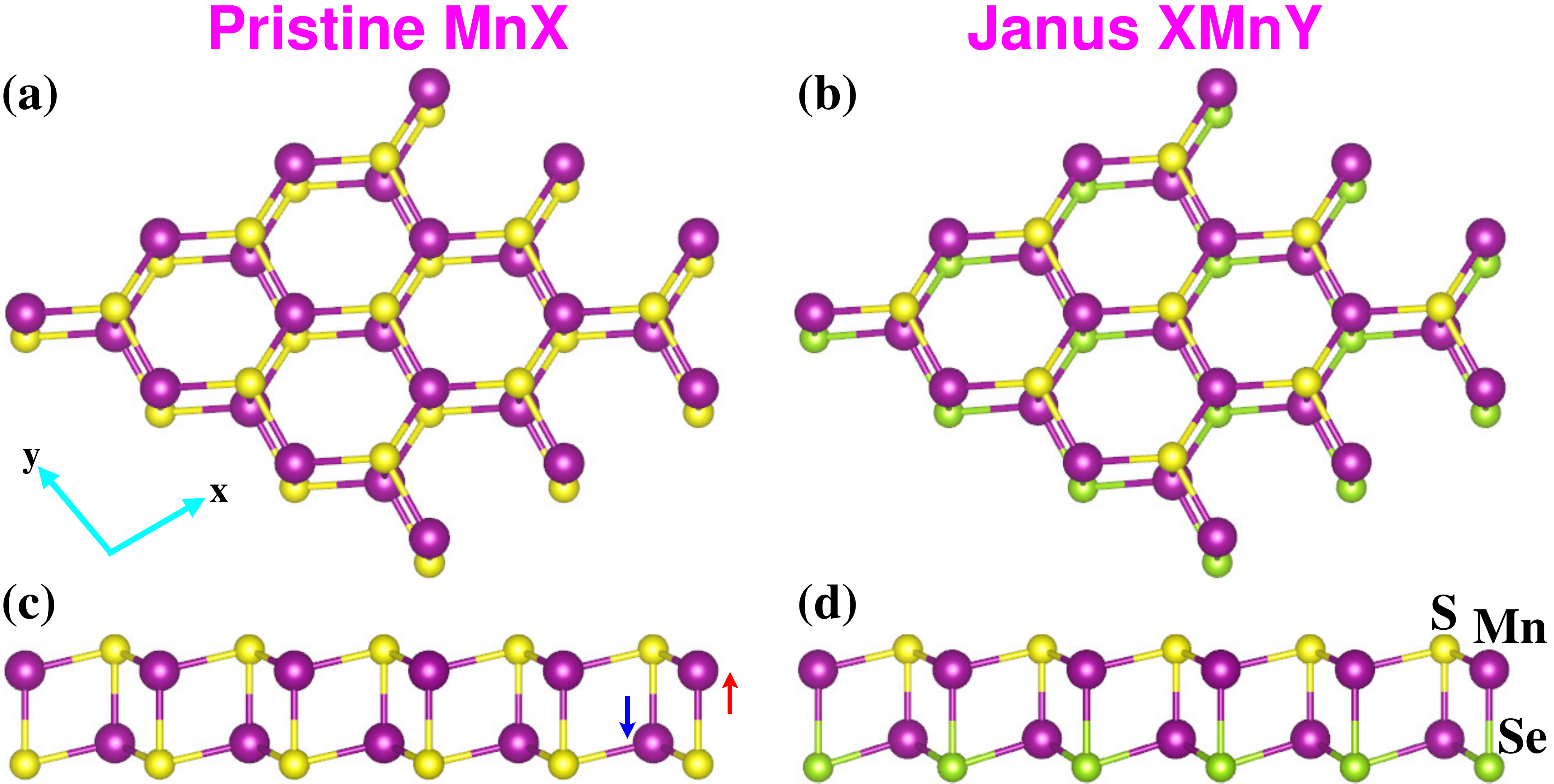}
\caption{(a,b) Top views, and (c,d) Side views of pristine monolayer MnX and Janus XMnY (X, Y= S, Se, Te), respectively. Here sulphur (S), selenium (Se) and manganese (Mn) atoms are shown in yellow, green and purple colors.}
\label{fig:fig1}
\end{figure}

\section{Computational Details}

We performed first-principles calculations using density functional theory (DFT) and projector augmented waves method\,\cite{paw1,paw2} as implemented in the Vienna \textit{ab-initio} simulation package (VASP) \cite{vasp}. The generalized gradient approximation was used in the Perdew-Burke-Ernzerhof parametrization to describe exchange-correlation effects. The plane wave cutoff energy was set to a large value of 450 eV. For structural relaxations, a gamma-centered $12\times 12\times 1$ k-mesh was employed whereas in the self-consistent calculations, Brillouin zone integration was as performed using a dense $16\times16\times 1$ k-mesh. For the calculations of density-of-states (DOS), we used tetrahedron method with Bl\"{o}chl corrections using a more refined $20\times20\times 1$ k-mesh. Owing to their significance, spin-polarization and SOC were especially included in structural relaxation, band structures and DOS calculations. Furthermore, to provide an accurate description of strongly correlated Mn d-orbitals, we made use of the Hubbard $U$ correction for strongly-correlated systems and set an onsite Coulomb (U) and exchange (J) parameters to the values of 2.3 eV and 0 eV, respectively, as used in a previous study for monolayer MnSe \cite{aapro2021synthesis}. For the computation of phonon dispersion across the irreducible Brillouin zone, we utilized the finite-displacement method implemented in Phonopy package \cite{togo2015first} and used a $4\times 4\times 1$ supercell of MnX or XMnY (X, Y = S, Se, Te) sampled using a gamma-centered $2\times 2\times 1$ k-mesh. To confirm thermal stability, \textit{ab-initio} molecular dynamics simulations were performed using the Nos\'{e}-Hoover thermostat (canonical ensemble) at 300 K for a time interval of 2 ps (time step of 1 fs). For the magnetocrystalline anisotropy, SOC is taken into account non-self-consistently after an accurate collinear calculation. In calculating the iterative solution of Kohn-Sham equations, we achieved an energy convergence of $10^{-6}$ eV and a force convergence of $10^{-3}$ eV/\AA. We have also used a 15\,\AA\,thick layer of vacuum in the out-of-plane direction to avoid periodic image interactions. Finally, vaspvis \cite{yu2021dependence} and pyprocar \cite{pyprocar} packages were employed for pre- and post-processing and plotting of data.

\section{Results and Discussion}
\subsection{Structure and structural stability}

Figure \ref{fig:fig1}(a-d) shows top and side views of the pristine monolayer MnX and the Janus XMnY structure (X, Y= S, Se, Te), respectively. Both AFM families adopt the trigonal crystal structure with space group \textit{P$\Bar{3}$m}1 (no. 156) which resembles that of $AA$-stacked bilayer silicene discussed in Ref. \cite{sattar2018stacking}. Under a static and dynamic point group symmetry of $C_{3v}$, both inversion and time-reversal symmetry are absent in the pristine and Janus materials. Looking first at the details of the unusual atomic structure of these manganese chalcogenides, we note that each of the Mn atoms within the top and bottom $xy$-planes is attached to three neighboring chalcogen X (X = S, Se, Te) atoms (see Figure \ref{fig:fig1}(a)) with a characteristic buckling height of 0.56\,\AA,\,0.73\,\AA,\,and 0.97\,\AA\,, respectively. For the unit cell crystal structure consisting of two Mn and two chalcogen atoms, the in-plane bond lengths (Mn$-$S: 2.44\,\AA\,and Mn$-$Se: 2.56\,\AA) is slightly shorter than the out-of-plane values (2.48\,\AA\,and 2.59\,\AA), respectively, depicting stronger in-plane intralayer interactions. Monolayer MnTe however experiences equal bond lengths in both in-plane and out-of-plane directions (Mn$-$Te: 2.77\,\AA). For the case of Janus monolayer XMnY (X, Y= S, Se, Te) shown in Figure \ref{fig:fig1}(b,d), different bond lengths and buckling are an obvious consequence of the dissimilar chalcogen (X, Y) atoms on the top and bottom surface. Considering SMnSe, SeMnTe, and SMnTe as three possible Janus structures, we observe non-equivalent buckling heights in the top and bottom $xy$-planes having magnitudes of 0.56\,\AA/0.76\,\AA, 1.01\,\AA/0.72\,\AA, and 0.57\,\AA/1.06\,\AA\,between S-Mn/Se-Mn, Se-Mn/Te-Mn, and S-Mn/Te-Mn atoms, respectively. Surprisingly, the net magnetic moment (for pristine as well as Janus manganese chalcogenides) remains zero, signaling the presence of AFM order, as discussed in the next section. The optimized lattice parameters for all studied cases are listed in Table \ref{table:table1}, whereas structural data are provided in the supplemental material.   

\begin{figure}[!t]
\includegraphics[width=0.47\textwidth]{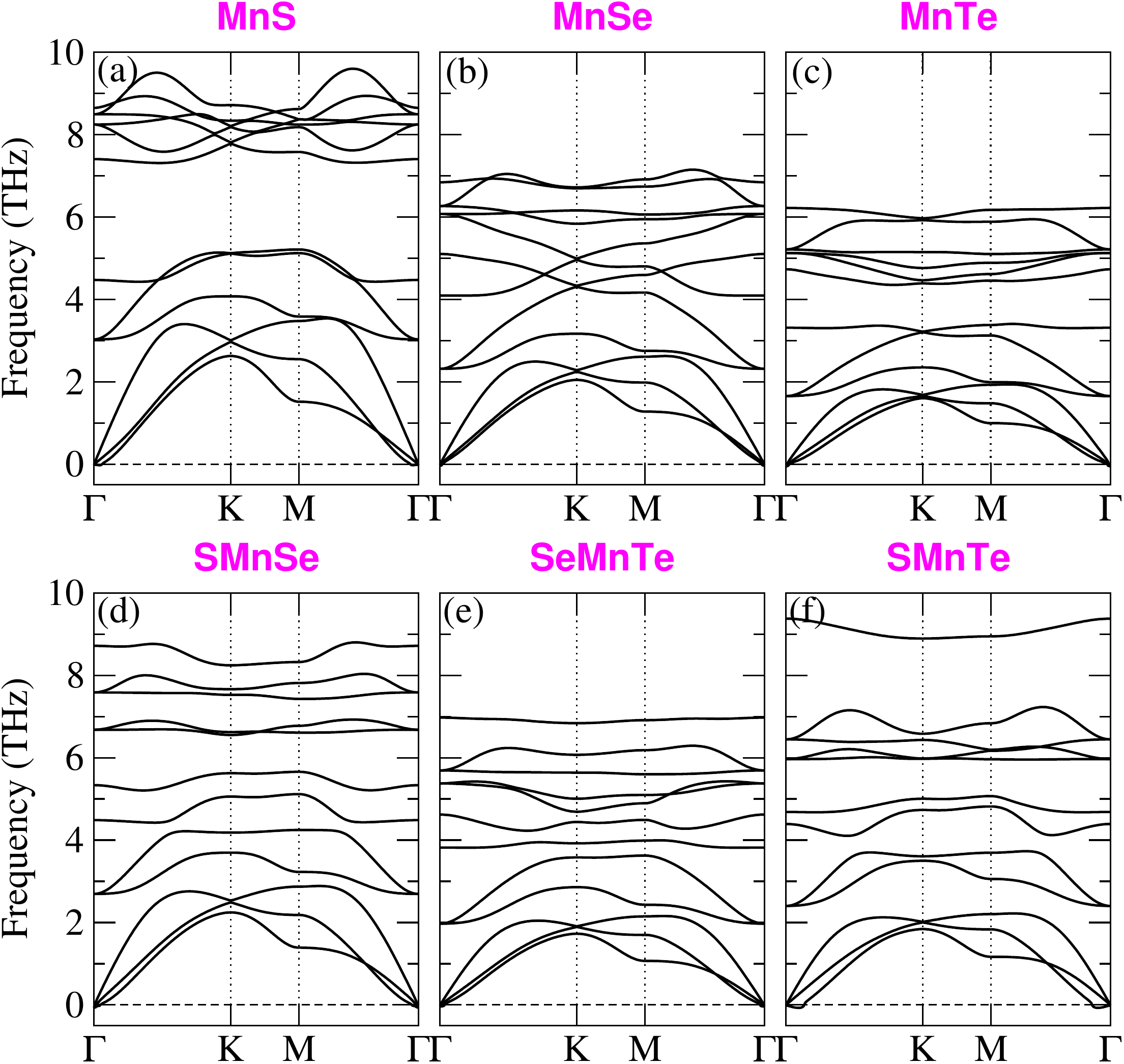}
\caption{(a-c) Phonon band structures of pristine monolayer MnX, and (d-f) Janus XMnY (X, Y= S, Se, Te), respectively. The lowest three bands correspond to acoustic phonons.}
\label{fig:fig2}
\end{figure}

We next investigate the dynamic stability of AFM monolayer manganese chalcogenides by calculating phonon dispersion along the high-symmetry lines in a typical hexagonal Brillouin zone ($\Gamma\rightarrow\text{K}\rightarrow\text{M}\rightarrow\Gamma$) and the thermal stability by performing \textit{ab-initio} molecular dynamics simulations. As shown in Figure \ref{fig:fig2}(a-f), all phonon modes display positive frequencies across the momentum space indicating the stability of each structure. Because heat conduction is mainly governed by acoustic phonons (first three modes), it is worth pointing out the decrease in the cutoff frequency of the acoustic modes with the increase of the atomic mass from MnS to MnTe (see Figure \ref{fig:fig2}(a-c)). Another interesting feature is the absence of coupling between acoustic and optical modes in the phonon band structures of pristine MnX which possibly hints at low phonon scattering rates. However, refer to the case of Janus materials in Figure \ref{fig:fig2}(d-f), the gap between acoustic and optical modes increases which indicates rigid bonding and large phonon lifetime. On the other hand, we also checked the thermal stability of these monolayer systems, and the results are presented in the supplemental material (see Figure S1). Since thermally-induced energy fluctuations are within a small energy window of 10 meV and ensuing minimal lattice distortions do not change the crystal structure, all investigated structures are stable at ambient conditions. 

\begin{table}[!b]
\caption{Optimized lattice parameters (a,b), Mn1-X-Mn2 bond angle (for Janus materials bond angles are different for the top and the bottom surfaces), exchange energy per Mn atom $\Delta$E$_{ex}$ = (E$_{FM}$-E$_{AFM}$)/2, total magnetic moment of top and bottom Mn atoms (Mn1/Mn2), band gaps (without SOC) alongside the position of VBM and CBM in the Brillouin zone, and magnetocrystalline anisotropy energy (MAE) per Mn atom.} 
\begin{tabular}{|c|c|c|c|c|c|c|}\hline
Material            & MnS     & MnSe   & MnTe    & SMnSe    & SMnTe    & SeMnTe           \\ \hline
a = b (\AA)         & 4.11    & 4.28   & 4.52    & 4.19     & 4.30     & 4.40             \\ \hline
Mn1-X-Mn2             & 78$^o$  & 74$^o$ & 70$^o$  & 77$^o$,  & 74$^o$,  & 32$^o$,          \\ 
bond angle          &         &        &         & 73$^o$   & 68$^o$   & 67$^o$           \\ \hline
$\Delta$E$_{ex}$   & 0.250   & 0.235  & 0.225   & 0.245    & 0.255    & 0.230           \\
   (eV)             &         &        &         &          &          &                  \\ \hline
Mag. mom.                & 4.36    & 4.38   & 4.37    & 4.34     & 4.29     & 4.33             \\
($\mu_B$)      & 4.36    & 4.38   & 4.37    & 4.38     & 4.38     & 4.39             \\ \hline
Band gap            & 1.78    & 1.76   & 1.29    & 1.40     & 0.78     & 1.05             \\
(eV)                & $\Gamma \rightarrow K$ & $\Gamma \rightarrow M$  & $\Gamma \rightarrow M$ & $\Gamma \rightarrow K$ 
                    & $\Gamma \rightarrow K$ & $\Gamma \rightarrow M$                       \\ \hline    
 MAE                & -0.25   & -0.41  & -0.65   & -0.13    & -0.27    & -0.38             \\ \cline{2-7}
 (meV)              & \multicolumn{6}{c|}{In-plane easy axis}                              \\ \hline
\end{tabular}
\label{table:table1}
\end{table}

\subsection{Electronic and magnetic properties}

\begin{figure}[!t]
\includegraphics[width=0.48\textwidth]{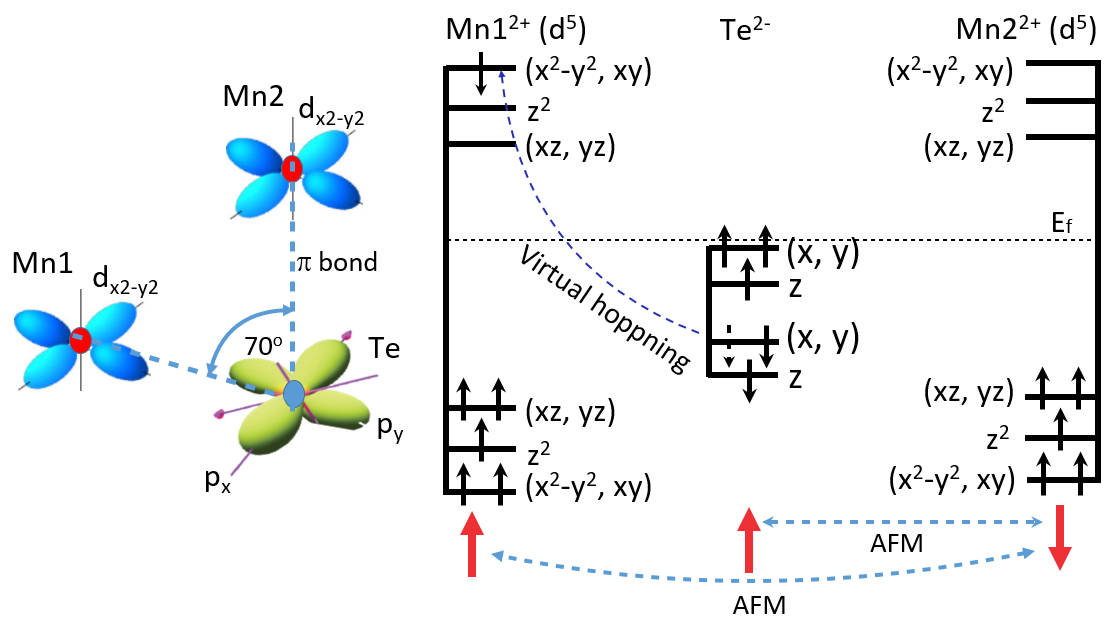}
\caption{Mechanism of AFM coupling between two Mn ions with $d^5$ configuration in MnTe resulting from superexchange via $p$ orbitals of Te ion. 
}
\label{AFM}
\end{figure}

We now discuss the electronic and magnetic properties. Monolayer MnX and Janus XMnY constitute a family of robust AFM materials with a much lower energy than their FM state as listed in Table \ref{table:table1}. Taking $\Delta$E$_{ex}$ to be the energy difference per Mn atom, we find this to be as large as 0.225 eV for monolayer MnTe. In order to explain the origin of the anti-ferromagnetic order in these materials, we consider the nature of coupling in the Mn1-Te-Mn2 bond in MnTe, in which Mn$^{2+}$ cations have the $d^5$ configuration. The exchange and crystal field ($C_{3v}$) splittings of the $d-$orbitals of Mn$2^+$ cation and the $p-$orbitals of the Te$^{2-}$ anion are shown in Figure~\ref{AFM}. Since all five Mn $d_{up}$ states are fully occupied, charge transfer can occur through virtual hopping only from $p_{dn}$ states of Te to $d_{dn}$ states of Mn. Due to the symmetry constraint, ($p_x, p_y$) states can hybridize only with ($d_{xy}$, $d_{x^2-y^2}$) states as indicated by the dotted blue line. The Te $p-$orbital now becomes spin-polarized aligned parallel to Mn1 ion as shown by the red up-arrow, and interact with the Mn2 ion. Since Mn2 lies just at the top of Te, the ($p_x, p_y$) and ($d_{xy}$, $d_{x^2-y^2}$) states can form a $\pi$-bond as shown on  the left of Figure~\ref{AFM}. According to Goodenough-Kanamori rules \cite{goodenough1955theory,kanamori1960crystal}, such bonding leads to an AFM coupling between Te and Mn2 ions. Consequently, occupied $d-$states of Mn2 ion flip the spin as indicated by the red down-arrow. Therefore, Mn1 and Mn2 couple anti-ferromagnetically via the superexchange mechanism. We note that the superexchange can also be mediated via Mn $d_{z^2}$ and Te $p_z$ states, which is permitted by the symmetry. However, in this case, the states form a $\sigma$-bond between Te and Mn2.

\begin{figure}[!b]
\includegraphics[width=0.475\textwidth]{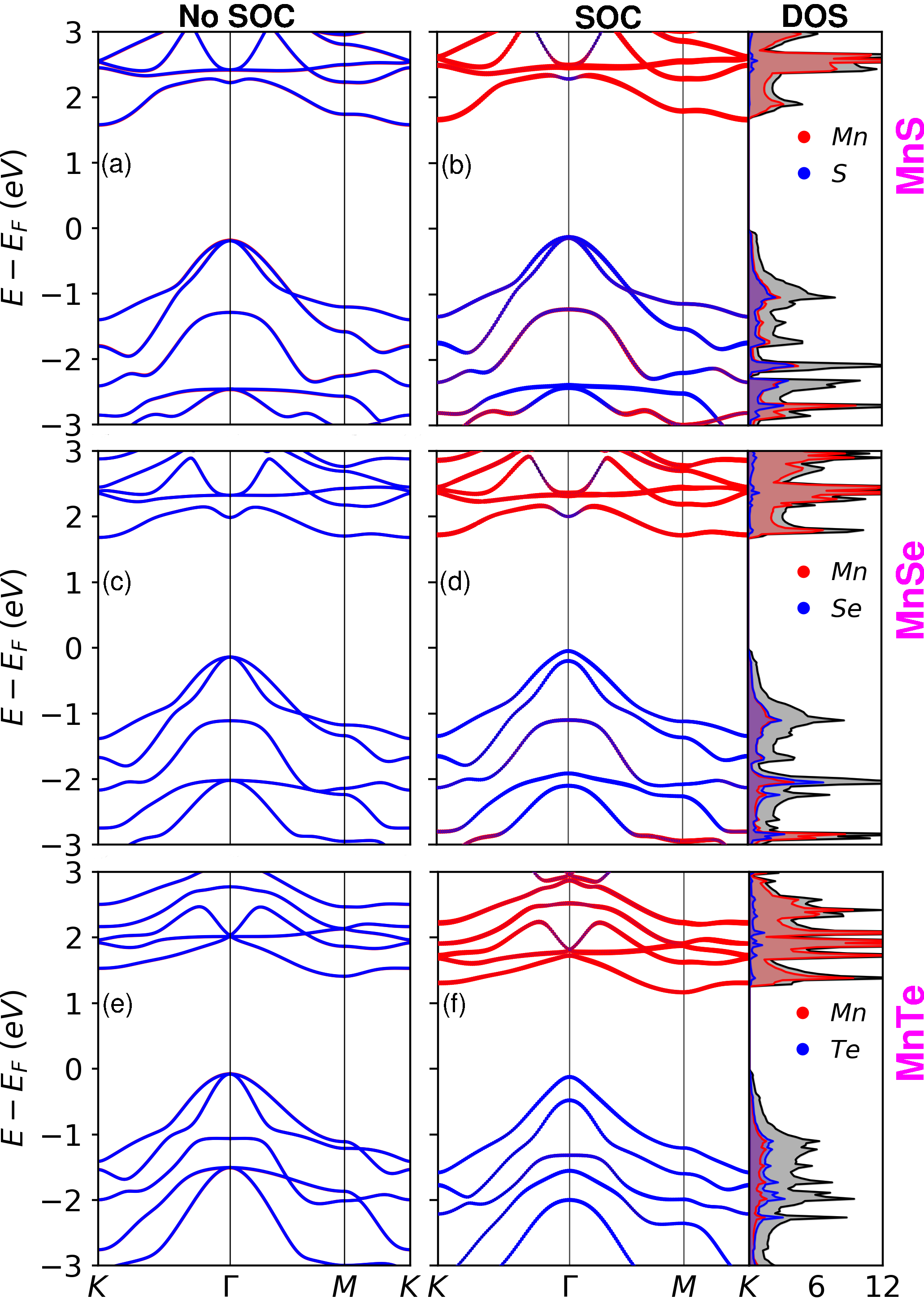}
\caption{(a,c,e) Electronic band structures of the AFM monolayer MnX (X= S, Se, Te) without SOC. (b,d,f) Element-projected band structures and DOS plots incorporating SOC effects.}
\label{fig:fig3}
\end{figure}

In their ground state, each manganese atom in the top/bottom planes of MnX (i.e., Mn1/Mn2) has a magnetic moment of $\sim 4.4\,\mu_B$ but pointing in opposite directions in the two planes, resulting in a net zero magnetic moment. In Janus XMnY, despite a difference of magnetic moment between Mn1/Mn2 (e.g., 4.33$\,\mu_B$/4.39$\,\mu_B$ for SeMnTe), the overall magnetic moment remains zero due to charge transfer and redistribution between the constituent atoms of top and bottom planes. This effect is largest for SMnTe owing to the electronegativity difference between S (2.58), Mn (1.55), and Te (2.1) atoms. In addition to the opposite magnetic moments of Mn atoms, the presence of dissimilar chalcogens in the AFM Janus monolayer chalcogenides gives rise to an asymmetrical potential in the out-of-plane direction responsible for lifting the spin-degeneracy of the bands, which is discussed in the subsequent section. 

We next discuss spin-polarized electronic band structures (without and with SOC effects) and DOS plots of AFM monolayer MnX as shown in Figure \ref{fig:fig3}(a-f). Pristine MnX turns out to be a semiconductor with indirect band gap of 1.78 eV, 1.76 eV, and 1.29 eV for monolayer MnS, MnSe, and MnTe, respectively. We notice that the energy difference between conduction band minima's (CBM) at the high-symmetry K-point and M-point is small (especially for MnS and MnSe), whereas the valence band maxima (VBM) is present at the $\Gamma$-point. Hence, the energy difference between highest occupied (valence) and lowest unoccupied (conduction) bands is used to evaluate band-gap values, which alongside their position in the Brillouin zone are listed in Table \ref{table:table1}.  We note from the bandstructure that, although both inversion and time-reversal symmetries are broken in these systems, each band is spin degenerate due to the AFM order (spin-up bands of the top layer Mn atom coincide with the spin-down bands of the bottom layer Mn atom). The valence bands are predominantly of $p$ character of the chalcogen atoms whereas the conduction bands are formed predominantly by the $d$ orbitals of Mn atoms. At the $\Gamma$-point, the VBM corresponds to the two-fold degenerate irreducible $E$ representation of the $C_{3v}$ point group formed by ($p_x, p_y$) orbitals. Away from the $\Gamma$-point the orbital degeneracy is lifted due to the reduction of the symmetry.

In presence of the SOC, the orbital degeneracy at the $\Gamma$-point is lifted as shown in Figure \ref{fig:fig3}(b,d,f). We observe the largest band-splittings of 350 meV for the monolayer MnTe owing to the substantial SOC strength of Te atoms ($\sim$ 0.5 eV) compared to Se ($\sim$ 0.22 eV) and S ($\sim$ 0.05 eV) atoms. 
%
%

\begin{figure}[!t]
\includegraphics[width=0.475\textwidth]{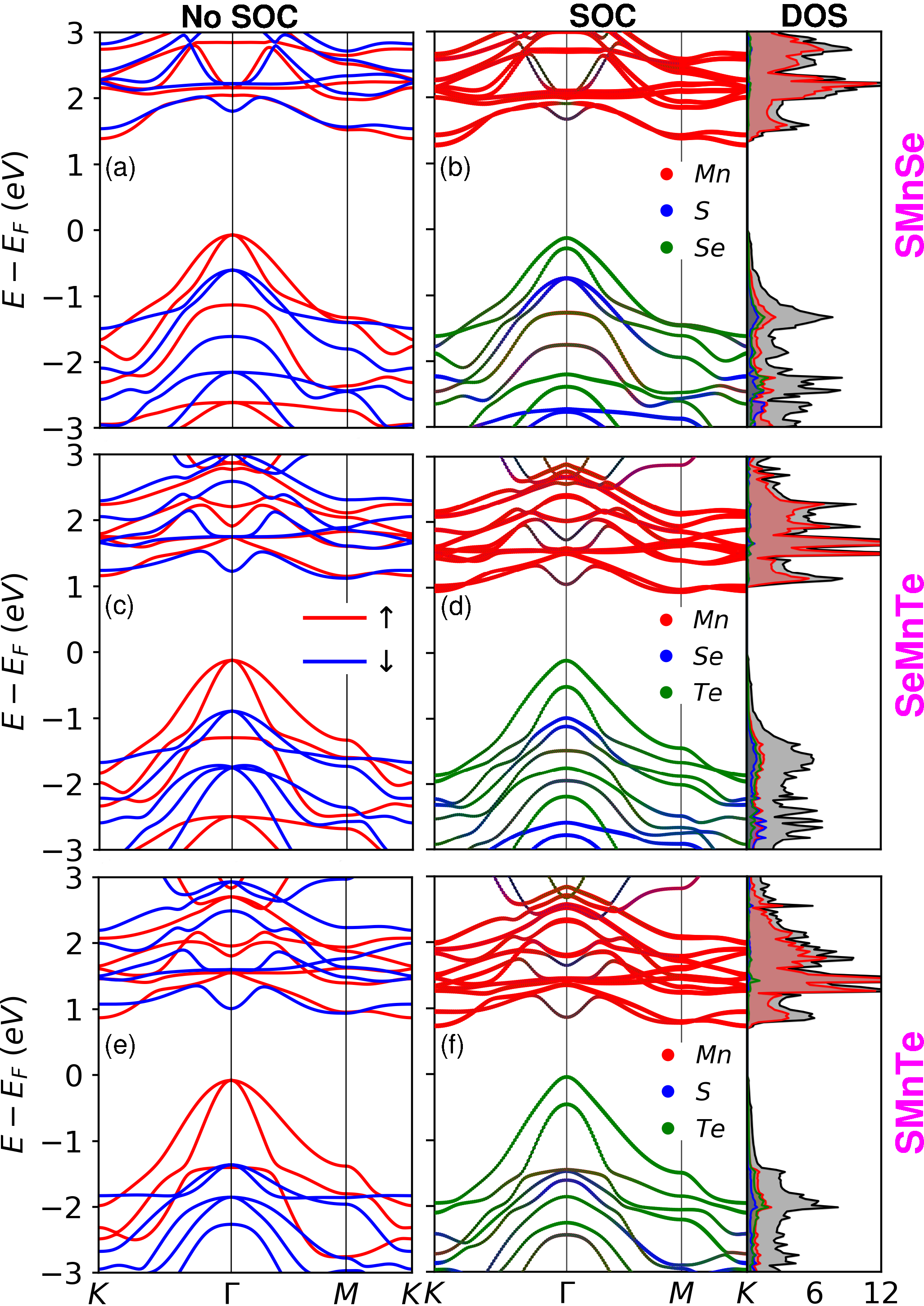}
\caption{(a,c,e) Electronic band structures of AFM Janus monolayer XMnY (X, Y= S, Se, Te) without SOC. (b, d, f) Element-projected band structures and density of states (DOS) plots of the same systems incorporating SOC effects.}
\label{fig:fig4}
\end{figure}

The orbital nature is further confirmed by plotting the total DOS (in black), and the element-projected DOS of Mn and chalcogen atoms (in red and blue colors, respectively). It is evident that at the neighborhood of the VBM, bands formed predominantly by chalcogen $p$ states and are hybridized with Mn $d-$states. The strength of the hybridization is strongest in MnS as it has the smallest lattice constant, whereas MnTe has the weakest hybridization. Due to the symmetry constrain of the $C_{3v}$ point group, ($p_x, p_y$) orbitals of chalcogen hybridize with ($d_{xy}$, $d_{x^2-y^2}$) orbitals of Mn as both pairs form the $E$ representation of the group whereas the $p_z$ orbital hybridizes with $d_{z^2}$ as both belongs to the $A_1$ representation. On the other hand, the ($d_{xz}$, $d_{yz}$) orbitals do not hybridize with any $p$ orbitals due to the orthogonality constraint. At the vicinity of the CBM, bands are essentially formed by the Mn $d-$states. Owing to band gaps of monolayer MnS and MnSe falling in the visible energy range of photons, it is anticipated that these materials will be useful for AFM opto-spintronics. The presence of flat bands and multiple valleys in MnX across different paths of Brillouin zone (namely $\Gamma \rightarrow$ K and $\Gamma \rightarrow$ M directions) offers great opportunities for valleytronics and AFM spintronics.

It is pertinent to mention that removing the constraint of strong-correlation on Mn $d-$electrons (i.e., no Hubbard U correction) in our calculations significantly reduces the energy gap. As an example, MnTe shows a band gap of 0.66 eV (see supplemental material, Figure S2) without the inclusion of an onsite Coulomb exchange (U) parameter, which demonstrates its importance of correlations in capturing correct band gap values.

We now proceed to the analysis of the electronic properties of the AFM Janus monolayer XMnY, where Mn is bonded to dissimilar chalcogens (X, Y= S, Se, Te). As in the pristine case, FM top and bottom $xy-$planes couple anti-ferromagnetically resulting in a net zero magnetization. The electronic band structures and DOS for these systems are shown in Figure \ref{fig:fig4}(a-f) without and with SOC effects. Janus manganese chalcogenides also show semiconducting behavior alike the pristine MnX discussed above. However, dissimilar chalcogen atoms in the Janus structures create an asymmetrical potential, 
which is responsible for the lifting of the spin-degeneracy between spin-up and spin-down channels as shown in Figure \ref{fig:fig4}(a,c,e). The band gap for monolayer SMnSe, SeMnTe, and SMnTe without SOC effects listed in Table \ref{table:table1} corresponds to the difference in energy between the CBM and VBM (spin-up bands shown in Figure \ref{fig:fig4}(a,c,e)). We also note from the figure that the top-most spin-up (red) and spin-down (blue) spin-channels in the valence band are separated by an energy of 0.52 eV, 1.01 eV, and 1.27 eV, respectively. Moreover, the inclusion of SOC results in band-splittings at the high-symmetry $\Gamma$-point as shown in Figure \ref{fig:fig4}(b,d,f) which further increases the gap between spin-up and spin-down channels. 

\begin{figure}[!t]
\includegraphics[width=0.475\textwidth]{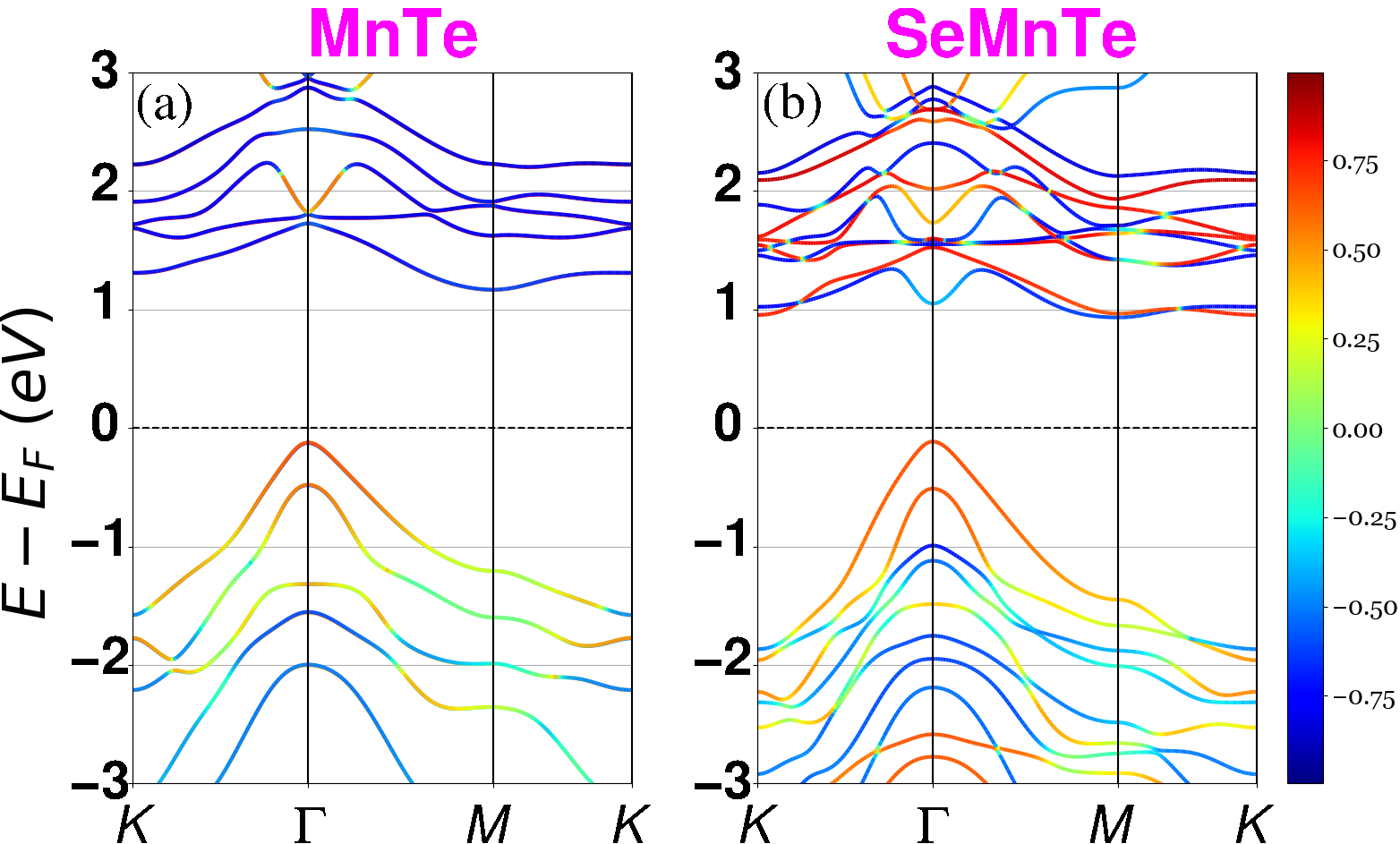}
\caption{Spin-projected band structure of (a) MnTe and (b) Janus SeMnTe. Red and blue color represent spin-up and spin-down channels, respectively.}
\label{fig:fig5}
\end{figure}

In order to clarify the nature of band gap and spin-orientation in both pristine and Janus manganese chalcogenides, we plot spin-projected band structures of monolayer MnTe and SeMnTe projected on $S_z$ component as shown in Figure \ref{fig:fig5}(a-b). Since due to the AFM order the spin-up and the spin-down channels are degenerate, the band gap is the same (1.29 eV) for both spin channels in the pristine MnTe (see Figure \ref{fig:fig5}(a)). However, in the case of Janus SeMnTe, shown in Figure \ref{fig:fig5}(b), the energy gap of spin-down channel is $\sim 1$ eV larger than that of spin-up due to staggered potential in the perpendicular direction.


\subsection{Magnetocrystalline anisotropy}

In order to establish the orientation of the magnetic staggered order-parameter, we have calculated the magnetocrystalline anisotropy energy (MAE) of all six materials, which is listed in Table~\ref{table:table1}. Our calculations show that the MAE is of the order of 1 meV and the easy axes are in-plane for all these materials. To explain the direction of easy axis, we consider the second order perturbation theory with SOC Hamiltonian, $H_{so}=\lambda L\cdot S$, as perturbation. Here $\lambda$ is SOC constant. To find the easy axis it is convenient to express the spin-orbit Hamiltonian with explicit dependence on the direction of the spin moment with respect to the coordinate axes. To derive the angular dependence we express the Hamiltonian in terms of ladder operators,
\begin{eqnarray}
H_{SO} = \lambda (L_xS_x + L_yS_y + L_zS_z) \nonumber \\ 
       = \lambda (L_zS_z + \frac{1}{2}L_+S_- + \frac{1}{2}L_-S_+) 
\label{hso}
\end{eqnarray}

Now let n($\theta$) be the magnetization direction where $\theta$ is the polar angle. Here we assume that the SOC energy does not depend on the azimuthal angle ($\phi=0$). Then the Hamiltonian can be expressed as 

\begin{equation}
\begin{split}
H_{SO} & = \lambda S_z(L_z\mathrm{cos\theta}+\frac{1}{2}L_+\mathrm{sin\theta}+\frac{1}{2}L_-\mathrm{sin\theta}) \\
      & +\frac{\lambda}{2}S_+(-L_z\mathrm{sin\theta} - L_+\mathrm{sin^2\frac{\theta}{2}}+L_-\mathrm{cos^2\frac{\theta}{2}}) \\
      & +\frac{\lambda}{2}S_-(-L_z\mathrm{sin\theta} + L_+\mathrm{cos^2\frac{\theta}{2}}-L_-\mathrm{sin^2\frac{\theta}{2}})  
\end{split}
\label{hmae}
\end{equation}

The dominant contribution to the energy that depends on the magnetization direction comes from the mixing between the highest occupied magnetic state, $|n>$, and the lowest unoccupied magnetic state, $|n'>$. Therefore, according to the second order perturbation theory, the ensuing lowering of the ground state energy by spin-orbit interaction is
\begin{equation}
\Delta E_{SO} = -\frac{|<n|H_{SO}|n'>|^2}{\epsilon_n-\epsilon_{n'}}
\label{deltaE}
\end{equation}

\begin{figure}[!ht]
\includegraphics[width=0.47\textwidth]{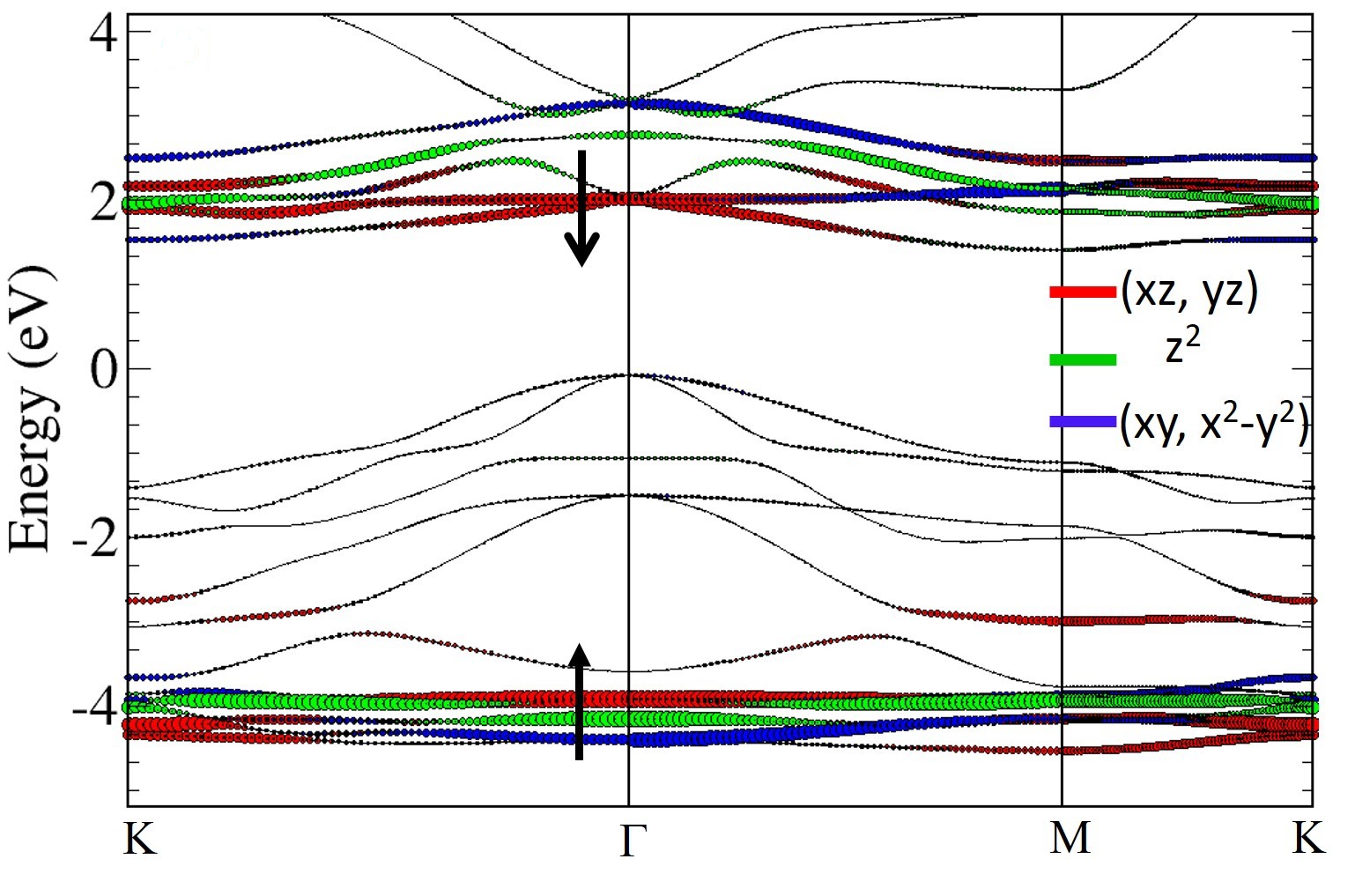}
\caption{Orbital ordering of Mn $d$ states of MnTe without SOC.}
\label{d_band}
\end{figure}

In order to understand the mechanism leading to the appearance of a preferential direction for the staggered magnetization, we consider the case of MnTe as an example. We have closely checked the orbital ordering of MnTe as shown in Figure~\ref{d_band}. It is evident from the figure that, in the $C_{3v}$ crystal field, the $d$-states of Mn in MnTe split into two 2-fold degenerate states at the $\Gamma$-point: ($d_{xy}, d_{x^2-y^2}$) being the lowest energy states and the ($d_{xz}, d_{yz}$) being the highest energy states for the occupied bands. The energy of the remaining 1-fold $d_{z^2}$ state lies between the degenerate states. The orbital ordering reverses in the unoccupied bands. We also note from the figure that the occupied and the unoccupied states have opposite spin states. 

Since the $|n'>$ state has $|\downarrow>$ spin state, only the $S_+$ term in Eq.~\ref{hmae} will couple the occupied and unoccupied states, which are listed in Table~\ref{mae_states}. Furthermore, since both occupied and unoccupied states are of the same orbital character $Y_1^{\pm 1}$, only the $L_z$ term of the Hamiltonian in Eq.~\ref{hmae} will give a non-zero contribution to the matrix element in Eq.~\ref{deltaE}. 
\begin{table}[!ht]
\caption{The highest occupied $|n>$ and the lowest unoccupied $|n'>$ spin-orbitals that determine the easy axis.} 
\begin{tabular}{|c|c|c|}   \hline
Material & $|n>$                                       & $|n'>$                                          \\ \hline
         & $(xz, yz)$                                  & $(xz, yz)$                                     \\ \cline{2-3}
MnTe     & $(i/\sqrt{2}[Y_2^{1}+Y_2^{-1}]|\uparrow>$,  & $(i/\sqrt{2}[Y_2^{1}+Y_2^{-1}]|\downarrow>,$    \\
         & $-1/\sqrt{2}[Y_2^{1}-Y_2^{-1}]|\uparrow>)$  & $ -1/\sqrt{2}[Y_2^{1}-Y_2^{-1]}|\downarrow>)$   \\ \hline
\end{tabular}
\label{mae_states}
\end{table}
Therefore, the lowering of the energy due to the spin-orbit interaction can be effectively expressed as   
\begin{equation}
\Delta E_{SO} = -\frac{|<n|\frac{\lambda}{2}S_+(-L_zsin\theta)|n'>|^2}{\epsilon_n-\epsilon_{n'}}
\label{eff_gs}
\end{equation}
The dependence of the ground state energy on the orientation of the magnetization axis is then, $\Delta E_{SO} \sim -sin^2\theta$, which is the lowest when $\theta = 90^o$, Therefore, the easy axis is in-plane.

\subsection{Effect of strain}

\begin{figure}[!ht]
\includegraphics[width=0.475\textwidth]{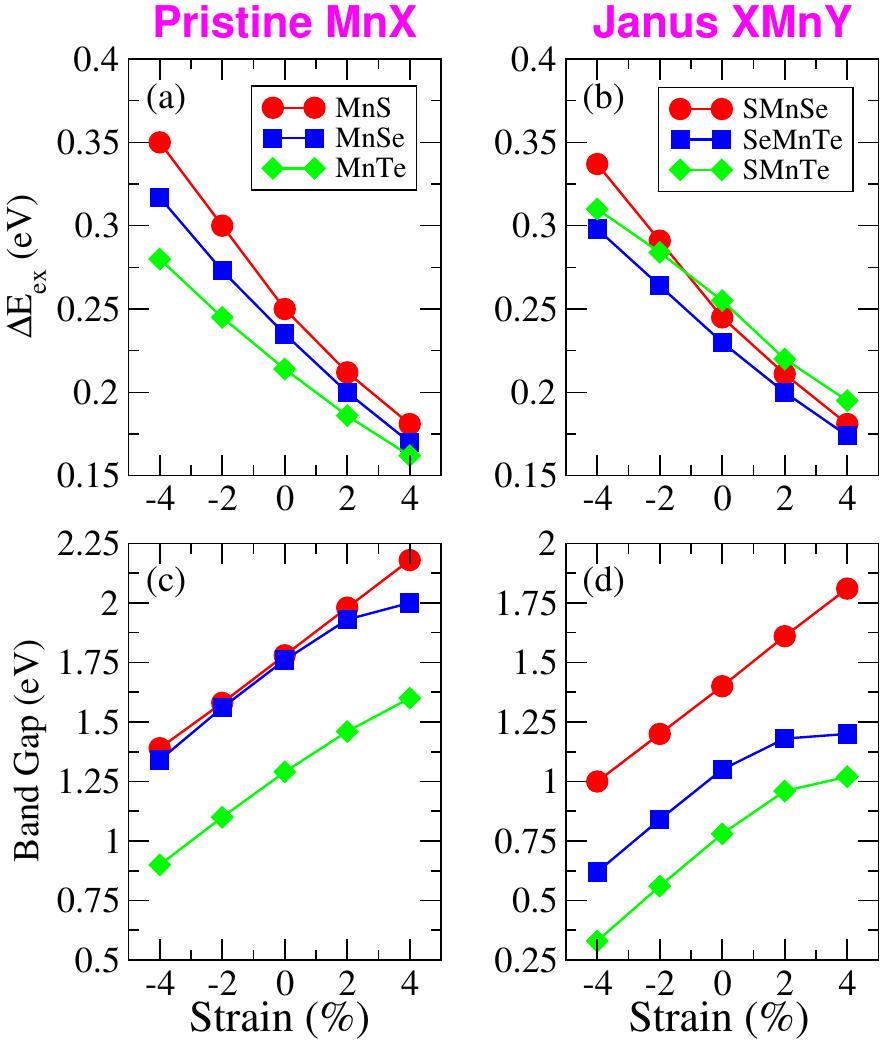}
\caption{Under biaxial tensile and compressive strain up to 4\%, (a,b) Energy difference per Mn atom between FM and AFM configurations, and (c,d) Magnitude of the electronic band gap of pristine and Janus manganese chalcogenides.}
\label{fig:fig6}
\end{figure}

Recent experiments have shown that strain control of the magnetic properties is possible, and can be used to induce a reversible AFM to FM phase transition in 2D AFM CrSBr at zero magnetic field \cite{cenker_reversible_2022}. It is interesting to check if such behavior can be also observed in this theoretical study. Therefore, we explore the effect of small biaxial tensile and a strain (up to $\pm$4\%) on the electronic and magnetic properties of the AFM manganese chalcogenides. It turns out that the ground-state remains AFM in all cases although the magnitude of exchange energy changes under the strain as shown in Figure \ref{fig:fig6}(a,b). At 4\% strain, the change $\Delta $E$_{ex}\sim0.35$ eV per Mn atom for MnS and SMnSe. 

The change in the exchange energy due to strain can be understood by considering the 2-site Heisenberg model $H=-2J_{ab}S_a\cdot S_b$, where a and b are atomic sites (Mn and Te in our case) with spins $S_a$ and $S_b$, respectively. The exchange coupling $J_{ab}$ can be expressed as      
\begin{equation}
J_{ab} = \frac{J_{ex}-CS^2}{1-S^4}
\label{strain}
\end{equation}
where $J_{ex}$, C, and S are exchange, Coulomb, and overlap integrals, respectively. The effect of the strain in a crystal is to change the interatomic distances between constituent atoms ($d_{Mn-X}$ and $d_{Mn1-Mn2}$ in our case), which decreases (increases) with negative (positive) strain, which in turn increases (decreases) the overlap integral S. Taking the example of MnTe, $d_{Mn-Te}$= 2.73\,\AA\, and $d_{Mn1-Mn2}$= 3.1\,\AA\ at -4\% strain, is smaller than non-strained values of 2.77\,\AA\,and 3.18\,\AA, respectively. We note from Figure~\ref{AFM} that such decrease in bond length increases overlap between $d$ and $p$ orbitals. It is now evident from Eq.~\ref{strain} that $J_{ab}$ increases (decreases) for the negative (positive) strain, which in turn increases (decreases) exchange energy. 


We have further investigated the effects of strain and the possibility to tune the electronic band gap. The results are presented in Figure \ref{fig:fig6}(c,d). Surprisingly, it is found that the band gap can also be controlled over a wide energy range by compressing or stretching these AFM systems. We observe a monotonic increase (decrease) in the energy gap for positive (negative) values of strain, which reaches up to 2.30 eV, 2 eV, and 1.6 eV against a tensile strain of +4\% for monolayer MnS, MnSe, and MnTe, respectively.

Our calculations show that the AFM order is quite robust under external strain in these materials. Covering a broad energy range in visible and infrared regime, these AFM materials provide interesting prospects of achieving both electrical (by external electric field) and optical control (using electromagnetic radiation).

\section{Summary and Outlook}

In summary, motivated by the recent experimental realization of AFM monolayer MnSe characterized by unusual atomic structure, we employed spin-polarized first-principles calculations to investigate the structural, electronic, and magnetic properties of the entire family of manganese chalcogenides, namely MnX and their Janus counterparts XMnY, with X and Y representing chalcogen atoms. Positive phonon frequencies across the Brillouin zone and minimal energy changes for small structural distortions in our molecular dynamics simulations confirmed the dynamical and thermal stability of all these AFM materials. In order to elucidate the onset of the AFM order and the large energy difference between the AFM and FM states, we discussed the role of $p-d$ orbital hybridization and interatomic bonding which suggests the presence of an underlying superexchange mechanism in these AFM ordered magnets. 
The finding of an indirect band gap which can be tuned over a wide energy range via biaxial tensile and compressive strain, suggests that both AFM pristine MnX and Janus XMnY chalcogenides are potential candidates for opto-electronics, opto-spintronics, neuromorphic computing, photovoltaics, and countless other applications. Similarly, a robust AFM behavior could be useful for magnetic memory operations due to the presence of inequivalent lattice sites. 

Another interesting outlook that should be investigated is electrical control of the electronic and magnetic properties, i.e., whether band gaps, spin splittings and MEAs can be tuned by the use of external electric fields. Aiming to achieve multiple functionalities, these layered AFM materials can be used to build van der Waals heterostructures, possibly to be integrated with non-magnetic as well as FM systems. Since this study is a preliminary investigation of the key characteristics of the whole family of ultrathin manganese chalcogenides, further theoretical and experimental studies are needed to probe and identify the practical aspects for various applications.   

\section{Acknowledgements}

Fruitful discussions with Hennu Pekka Komsa are greatly acknowledged. We thank Carl Tryggers Stiftelsen (CTS 20:71) for financial support. The computations were enabled by resources provided by the Swedish National Infrastructure for Computing (SNIC) at HPC2N and NSC partially funded by the Swedish Research Council through grant agreement no. 2018-05973.

\bibliographystyle{apsrev4-1}
\bibliography{main}

\end{document}